# Implications of EMU for the European Community

**CHRIS KIRRANE**

**Abstract**

Monetary integration has both costs and benefits. Europeans have a strong aversion to exchange rate instability. From this perspective, the EMS has shown its limits and full monetary union involving a single currency appears to be a necessity. This is the goal of the EMU project contained in the Maastricht Treaty. This paper examines the pertinent choices: independence of the Central Bank, budgetary discipline and economic policy coordination. Therefore, the implications of EMU for the economic policy of France will be examined. If the external force disappears, the public sector still cannot circumvent its solvency constraint. The instrument of national monetary policy will not be available so the absorption of asymmetric shocks will require greater wage flexibility and fiscal policy will play a greater role.

The paper includes three parts. The first concerns the economic foundations of monetary union and the costs it entails. The second is devoted to the institutional arrangements under the Treaty of Maastricht. The third examines the consequences of monetary union for the economy and the economic policy of France.

**Why EMU?**

Monetary union, which can be characterised by the combination of the freedom of capital movements and irrevocable exchange rates is one of the possible monetary regimes for economically highly integrated nations. A high degree of integration of the goods market, such as that seen in the countries of the European Community, does not automatically imply that choice. It is, in reality, the preference of Europeans for exchange rate stability that creates a link between the single market and monetary union. This preference has been repeatedly demonstrated since the abandonment of the Bretton Woods system through the establishment of the European currency snake and that of the European Monetary System.

It explains the micro-economic point of view (because the fixed exchange rate promotes economic integration), the macro-economic point of view (because it prevents non-cooperative attitudes), and also by institutional factors common policy management as in the CAP.[1]

Unlike economic integration theory that stresses unambiguous benefits, monetary union has costs (Mundell, (1961)). Economic analysis does not quantify the costs and benefits associated with the choice of exchange rate regime. At most it allows partial assessments (Emerson et al. (1991)). However, it provides guidance for establishing an objective balance of pros and cons to decide if this plan is or is not preferable to the others, particularly the current monetary regime of the majority of countries of the European Community: the EMS

---

[1] See, on this point, Giavazzi and Giovannini (1989).

**Advantages and disadvantages of monetary union**

The main economic reason for adopting a monetary union regime is micro-economic. The uncertainty about exchange rates is an obstacle to economic integration and the manifestation of welfare gains it provides. Historical observation shows that periods of floating exchange rates also correspond to periods of heavy unanticipated changes in real the exchange rate (Mussa (1986), Krugman (1989)). However, the direction of causality is certainly not absolutely established: the abandonment of the fixed exchange rate system in 1973 and the generalised floating of currencies are probably the result of accumulated tension in economies with nominal rigidities, the resolution increased the flexibility of exchange rates. But after twenty years of experience of floating exchange rates, there is little doubt that this diet leads to excessive volatility of exchange rates and the emergence of lasting misalignments.

Note however that micro-economic benefits of fixed exchange rates are not easily measurable. Thus, despite a lot of research on this issue, it is not empirically shown that the variability of nominal exchange rates led to a significant reduction in the volume of international trade (in the case of the EMS, see Sapir and Sekkat (1993)). As for the efficiency gains that the single currency could provide in promoting the integration of capital markets by eliminating distortions in the allocation of investments, stimulating competition and reducing the weight of uncertainty in investment decisions, they are hardly measurable (Baldwin (1991)).

The macro-economic benefits of monetary union belong, in turn, to strategic type considerations or, more broadly, to the political economy. It is well known that the choice of exchange rate regime is often inspired by such considerations (Argy (1989)), and this is verified with the EMS (Giavazzi and Pagano (1988)). In this perspective, monetary union is one of the ways of ensuring price stability, eliminating the risk of non-cooperative policy (see Kirrane [1993]), or promoting good management of public finances (Emerson et al. (1991)). In this matter, however, the costs and benefits depend on the alternative to monetary union and the quality of its constitutive principles. They are obviously not the same for Germany and Italy.

In the European case, monetary union should finally lead to the emergence of a European currency likely to compete with the dollar in its international currency functions. Economic gains related to the acquisition of such a status of international currency, however, are limited, at least in the medium term (Emerson et al (1991). Kenen (1993); Bénassy, Italianer and Pisani-Ferry (1993)). In the longer term, it is likely that the emergence of a multipolar currency world would be a factor of stability in a global economy where the economic weight of the United States is declining.

These potential advantages of monetary union must be weighed against the disadvantages resulting from the loss of autonomy in economic policy. It is indeed not possible to maintain an independent monetary policy if it is desired at the same time to maintain fixed exchange rates and the absence of hindrance to all capital movements (Padoa-Schioppa (1987)). Fiscal policy itself saw its margins reduced to the extent that monetary financing is banned, where the use of monetisation of debt is excluded, and where seigniorage is reduced (for the most inflationary countries) through the implied acceptance of a single inflation rate only compatible with fixed exchange rates. In the case of EMU, it is further limited by the principle that excessive deficits should be avoided.

It is known from Mundell (1961), the costs of this loss of autonomy depends on the distribution of disturbances affecting economies. As with any monetary rule, the choice of exchange rate regime is based on the nature and distribution of shocks that may jeopardise macroeconomic stability (Marston (1985)). In the case of monetary union, the key is whether the shocks are symmetric or asymmetric. In the first case, monetary union has only benefits because the fixed exchange rate plays as a substitute for cooperation and excludes non-cooperative behaviour. Participating countries all practice the same policy, which is also the desirable policy at the global level. However, if shocks are asymmetrical (by their origin or because of structural differences between countries), the lack of autonomy of monetary policy becomes a disadvantage. Debated somewhat abstractly by economists, this notion of 'asymmetric shock' found two striking illustrations with German unification and the savings behavior of British households. It was the magnitude of this asymmetry and the divergence of monetary policy responses that caused the UK, in September 1992, to make a forced exit from the EMS.

Many studies have been devoted to the identification of shocks within the European Community, particularly in a comparative perspective with the United States (see in particular Bayoumi and Masson (1991); Bayoumi and Eichen- Green (1992)). They usually lead to the conclusion that symmetric shocks dominate but that asymmetry is higher in the European Community and within the United States or Canada.

The dismantling of exchange rate controls raises risks to a system of fixed but adjustable exchange rates. To ensure exchange rate stability permanently, exchange rates must be irrevocably fixed. For this to happen, the EMS needs to have to a single monetary policy and thus a single currency regime.

**From EMS to Monetary union**

Monetary union should not be assessed in the abstract, but in reference to the EMS. Three main criticisms can be made about this regime.

First, the EMS has explicit goals of limited ambition (stabilisation of exchange rates, prohibition of competitive devaluations). Implicit rules have emerged during the period 1985-1992 (spacing realignments and incomplete correction of accumulated inflation differentials; see about Emerson et al (1991).) But their status is uncertain. The currency crisis of 1992-1993 could also lead to a change of system management principles, as proposed by the Monetary Committee of the Community (1993). Recent developments show that the 'new EMS' for years 1987-1992 (Giavazzi and Spaventa (1990)) had more dangers than previously believed. In fact, the EMS is primarily a monetary arrangement, and price stability is not one of the stated objectives of the system, although there is agreement that it played an important role in the decline inflation (Vissol (1990)). Now it is part of the objectives of EMU. Better integration of economic and monetary policy in EMU is expected to achieve the general economic objectives more effectively. Monetary policy should, in principle, take into account economic policies determined by the political authority [2] insofar as the objective of price stability is not jeopardised.

Secondly, even if it is rules-based and with balanced obligations, the EMS is characterised by asymmetry ever since the evolution of monetary policy gave the German mark the role of an anchor currency (around 1987 following the study of Weber (1990). Thus German monetary policy determines the key interest rates of other member countries and is mainly driven by exchange rate considerations. The hypothesis of German domination is certainly not confirmed by analysis of its strict version (Frattiani and von Hagen (1992)), probably because of capital controls, the flexibility provided by the bands of fluctuation and changes in risk premiums. Indirect evidence, however, leads one to consider that the Bundesbank plays a central role in the system.

---

[2] According to the Treaty on European Union, general economic policies must be decided by qualified majority by the Council, following the purposely rations of the European Council.

Now if this asymmetry could lend credibility to disinflation policies[3], it becomes sub-optimal from the time when many countries have achieved a satisfactory degree of price stability. The conduct of German monetary policy over the past two years shows clearly that where the occurrence of shocks affecting different countries unevenly, optimal monetary policy for a particular country (given the stance of fiscal policy) is suboptimal for the European Community as a whole. The role of an anchor currency-is also not without inconvenience to the Bundesbank itself whose monetary policy can be disrupted by the internationalisation of the mark (Köhler (1993)). The goal of the EMU is to establish a common monetary policy while retaining the benefit of price stability.

Finally, the events since September 1992 raise questions about the stability of the EMS. It is known from Obstfeld (1988) that speculative attacks were possible. Official reports (Padoa Schioppa (1987), Emerson et al. (1991)) had questioned the stability of a system of fixed exchange rates in a system with freedom of capital movements. But such attacks had not been observed. The currency crisis of 1992-1993, including the effect of the speculative attacks against the franc, forced a precise examination as to whether the EMS has been the theatre of such attacks. Research on these issues is just beginning (see, in particular, Eichengreen and Wyplosz (1993)). The reports of the monetary authorities blame the economic policies of the countries that had to devalue and had previously refused to make necessary adjustments to parities. The authorities now advocate a more flexible system (Monetary Committee (1993)). But if it were confirmed that the EMS can be subject to multiple equilibria, the merits of its continuation beyond a transitional period would be seriously in question.

**Is the single currency necessary?**

The Treaty on European Union states that after stopping the conversion of currencies the European Council 'shall take other measures necessary for the rapid introduction of the ECU as the single currency'. This vague wording opens the possibility of a phase III, during which the parities would be fixed and bands of fluctuation completely eliminated, but where different national monetary units continue to be in use. This precaution refers in fact to the technical time required for the introduction of the new currency but it raises questions about the interconnection of payment systems and about the cost/benefits of the single currency.

From a macroeconomic perspective, the irrevocable fixing of parities is equivalent to the introduction of a single currency. In particular, it strictly imposes the same degree of centralisation of monetary decisions. Micro-economically, the irrevocable fixing of exchange rates should lead to a transition to ECU interbank transactions and the 'wholesale' market, provided that the central banking system keeps its accounts in ECU and that governments denominate their debt in that currency (Kenen (1992)). One can well imagine the development of the single currency limited to the financial sector. However, there are differences between a monetary union simply based on fixed exchange rates and a monetary union with a single currency. In the case of maintaining separate currencies, the reduction of transaction costs related to monetary unification would not be fully exploited, the disciplines of the monetary union would be less visible to households and employees and finally, the credibility of the monetary union would be weaker (Emerson et al. (1991)).

---

[3] Note, however, that despite the rationalisation given by Giavazzi and Pagano (1988), applied studies have not been able to highlight the supposed benefits of disinflation in the EMS. See among recent works, Egebo and Englander (1992), or Blanchard and Mute (1993). De Grauwe (1989) even considers that the EMS has increased the cost of disinflation policies by preventing the use of the exchange rate and thereby imposing the gradualist strategies.

So there is an advantage, no doubt substantial, to go through with the logic of unification and to substitute national currencies with a single currency quickly. The introduction of the ECU, however, can done using either of two alternative strategies, gradualism and the 'big bang'.

### The big choices of the Treaty

Basically, the Maastricht Treaty is characterised by two major choices: to entrust the conduct of monetary policy to a single Central Bank, to guarantee its independence, and assign its aim as stability of prices; and that of preserving the autonomy of national budgetary policies while supporting them with devices designed to ensure budgetary discipline.

### The independence of the Central Bank

It is easy to decide on the principle of a Central Bank in any unified monetary space so that decisions are universally binding on all members. If several monetary authorities coexisted, each would have an incentive to increase monetary creation so as to increase revenues from seigniorage (Krugman (1989b)), or to promote businesses within its jurisdiction. Each member passes on to the others the costs of its inflationary policy and this quickly lead to a chaotic situation. The situation of the ruble zone provides a striking illustration (Big (1993)). The independence of the Central Bank is therefore necessary to maintain irrevocably fixed exchange rates but the choice to assign a priority objective of price stability is debatable.

In France and the UK, the tradition is that monetary policy is part of the range of policy instruments available to the authorities to achieve economic and social objectives that they set. In such a context, the Central Bank must remain in the service of general economic policy. In Germany and the Netherlands, the Central Bank, in contrast, is independent and specialised: it has the primary task to achieve a goal of price stability. Thus, the monetary instrument is assigned to a specific objective - the fight against inflation - and the institution that manages it is protected against pressures which could make it deviate from its purpose.

These two competing models of monetary organisation each have advantages and disadvantages. The first allows authorities to conduct discretionary policy using all the instruments of macroeconomic policy, without being trapped in a rigid assignment. Depending on the situation and the nature of tax instruments used, it is possible to vary the assignment, and thus for example to assign a fiscal policy target price and the monetary policy objective of growth and employment (for an illustration of the gains of the joint use of both instruments in a monetary union, and Villa Sterdyniak see (1993)). In a setting that Tinbergen, such flexibility is preferable. But as shown by Barro and Gordon (1983), the lack of specific objectives has a cost, because private agents expect the reactions of monetary authorities and adopt behaviors that can negate the effects on the real economy. This is why many countries, including France, assigned specific objectives to monetary policy. Assigning to the central bank a specific task and making it independent gives credibility to the objective of low inflation in the eyes of private agents and can overcome the disadvantages of discretionary policies. They learn, through experience, that the central bank is not ready to sacrifice this goal for the immediate benefits of supporting growth. The downside is, of course, that economic policy loses its flexibility and the possibilities of coordination between monetary and fiscal policies.

Economic theory and recent empirical research provides arguments for the independence of a central bank. Among others, the theoretical work of Kydland and Prescott (1977), Barro and Gordon (1983), and Cohen and Michel (1990) suggest that the cost of maintaining flexibility in the assignment of instruments is at a cost of a higher level of average inflation without gains in terms of growth. Applied available studies confirm the lack of positive correlation between inflation and growth in the medium term. They also indicate that, to the extent it can be quantified, the degree of independence of the central bank is negatively correlated with the level of inflation (Grilli, Masciandaro and Tabellini (1991) Alesina and Summers (1990) and Cukiermann, Webb and Neypati (1992)). In other words, the independence of the central bank results in lower inflation without a cost to growth. This valid statistical correlation over a long time obviously excludes either exceptions to the rule or temporary deviations, or the influence of other factors such as the degree of consensus around the goal of controlling inflation.

The economic developments of the past two decades indicate that these arguments should be taken seriously. The development and the deregulation of financial markets affect the conduct of monetary policy: the markets interpret the signals sent by the authorities and react to changes in the value of a particular instrument. Economic policy is thus less an engineering art in which the important thing is to have a range as wide as possible instruments. It is increasingly a game between policymakers and private agents, in which the clarity of objectives and continuity of decisions are conditions to the effectiveness of monetary policy.

In a monetary union, the choice of an independent central bank is needed more so, as shown in a series of articles by Laskar (1991a, 1991b and 1993). Indeed, the choice of exchange rate flexible, asymmetric and symmetric monetary union depends on the degree of independence of central banks. Several arguments are then used to argue that independence is in favour of monetary union:

-even in the event of complete convergence of savings and perfect symmetry of shocks, it may be that a flexible exchange system is preferred over a monetary union that does not have an independent Central Bank because of Rogoff's argument (1985)[4]. D. Laskar shows, however, that when a central bank is independent, a monetary union is always less advantageous than a flexible exchange rate system; a monetary union with an independent Central Bank is less inflationary which helps low inflation countries; moreover, as the specific shocks affecting each of the member countries tend to partially offset each other, the need to use monetary policy for the purpose of stabilising the economy in the face of shocks, is reduced. The European Central Bank can thus be more concerned with the general objective of price stability.

---

[4]The so-called 'counterproductive cooperation' argument. In this case, the monetary union is equivalent to a cooperation between central banks in flexible exchange rate system.

Finally, there is an institutional argument for a central bank independent. In a decentralised budgetary organisation, monetary policy assumes a greater share of the responsibility of the overall political economy. Federal states can respond to this situation by choosing an independent Central Bank.

The Statute of the European Central Bank (ECB) ensures that its independence will be at least as strong as that of Central Banks national authorities - including the Bundesbank[5]1. On the other hand, the action of the ECB will not immediately benefit from the same legitimacy with public opinion as that of Central Banks around which have a long-standing, solid consensus (Reeh (1993)).

Independence does not imply, however, (and must not imply) absence of dialogue and coordination. The Treaty and the Protocol on the Statute of the ECB specify its relations with the other European institutions. This obviously does not guarantee the quality of the dialogue, but at least corrects the image of splendid isolation of the future ECB that is sometimes given. Beyond procedures, a balance will have to be found between monetary and economic authorities in a practice reconciling respect for areas of competence and effective coordination. This is how the ECB can acquire its full legitimacy from national opinions.

**Budgetary discipline**

If the creation of a single, independent Central Bank brings together a broad consensus among economists, the question of whether to impose constraints on national fiscal policies is more controversial. The need to maintain the autonomy of national budgetary policies is hardly disputed. Macro-economically, it results from the need to compensate for the loss of the monetary instrument at the national level to cope with shocks of an asymmetrical nature, even though it is true that budgetary and monetary policy are perfectly substitutable. It is consistent with the chosen option of a Community budget of very small size and the maintenance of the Member States in most of the budgetary functions. Debate on the one hand, on budgetary discipline, i.e. the usefulness or otherwise of set limits on deficits and public debts and, on the other hand, on the degree of desirable coordination in a monetary union.

From an economic point of view, it is justified that excessive deficits be subject to Community multilateral surveillance procedures accompanied by a binding procedure based on safeguards (a) if an externality occurs between other members of the union, and (b) if the only mechanisms of the financial market are not sufficient to ensure compliance with budgetary discipline. These two points have been widely debated (for a recent overview, cf. Buiter, Corsetti and Roubini (1993)).

---

[5] The ECB will be independent of the Commission, Parliament, and governments. Its leaders will be appointed for eight years and will not be revocable. She will face twelve governments, which gives it greater de facto independence. And its status will be guaranteed by an international treaty, instead of national law.

The strongest argument for the existence of an externality is a solvency argument. It is based on the idea that a threat of default by a Member State would limit the ability of the Central Bank to conduct monetary policy of their choice. In a way, the Central Bank would be forced, if not to a bailout, at least to avoid precipitating the insolvency of a Member State by raising interest rates.[6] In that case, the other Member States would suffer from overly lax global monetary policy. And in case the country concerned would choose to withdraw from the union, the credibility of the union would be achieved (Emerson et al., (1991)). Some authors dispute that this possibility is likely (Buiter et al., (1993)). But it is difficult to exclude it.

Another argument, implicitly present in the treaty, is that a union's monetary policy could have a bias towards excessive deficits because of the dilution of the eviction effects and the relaxation of the equilibrium constraint outside, and apart from any problem of solvency, these deficits affect the overall savings-investment balance within the European Community. This argument is less robust for two reasons: incentives for budget management will be increased as the option of debt monetisation will be excluded; the savings-investment balance does not depend on public finances, but also private behaviour (the Italian case illustrates this clearly).

The argument in favour of organised discipline is based on the presumption of the limited ability of markets to properly value insolvency risks for sovereign debtors, particularly in a situation where the non-bailout commitment might lack credibility, and a doubt about the ability of governments to respond in an appropriate manner to market signals. A number of studies have attempted to appreciate this argument by analysing the situation of federations such as the United and Canada. The most precise (see, for example, Goldstein and Woglom (1991)) show a significant relationship between the level of debt of a State and the cost of its bonds. However, in this as in others, a doubt persists about the quality of the valuation of risk by the market (Artus (1990)), and the work of political economy is to question the reaction of governments to an increase in their cost of debt (Tanzi (1991)). If one of these arguments is true, it is justified to introduce constraints in order to avoid excessive deficits.

Two types of rules have been introduced in the Treaty: on the one hand, general rules prohibiting: 1) monetary financing, 2) privileged access by States to capital markets; 3) financial solidarity (no bail-out); on the other hand, the principle of exclusion of excessive deficits and the corresponding procedure, assorted quantitative criteria.[7] The criteria selected were subject to many critics (Buiter et al., (1993), Delessy et al., (1993), Masson et al. Melitz (1991)). In addition to their implications for the transition, the main problem is that it is not known exactly what they are aiming for. The text of the treaty oscillates between the notion of sustainability of the public debt (Article 109) and reference to the 'golden rule' (article 104 C3), and also takes into account the balance between savings and investment.

Beyond the debate on the theoretical basis, the criteria reflect principles of good management generally accepted in a steady state. But they present problems in a recession: the spontaneous evolution of public accounts over the period 1990-1993 indicates that the margin of 3% allowed for the public deficit is too narrow, unless one agrees that a deteriorated economic situation falls within the exceptions allowed by the treaty. They also pose a problem for the transition: the 60% public debt threshold is out of reach for several countries by the end of the decade, then some of them, like Belgium, have been living for almost ten years in a de facto monetary union.

---

[6]This risk seems higher than that of a budget bail-out.

[7] The purpose of this procedure is, in particular, to prevent the deficit of the public sector to permanently exceed 3% of GDP and its debt 60% of GDP.

The excessive deficit procedure is complex. It assumes in particular an assessment of the exceptional or permanent nature of the overrun criteria. Unlike the case of monetary policy, it implies a difference between countries, and this can lead to coalitions or bargains whose stakes go beyond the scope of fiscal policy. In fact, only practice will decide on the actual scope of this procedure. Its credibility will depend on its deterrence and its reaction to national opinions to a European Community recommendation or notice. Thus, its possible limits are probably less loopholes (valuation of assets) or procedural than to the political nature of the decision.

### The coordination of economic policies

In Europe, the coordination of economic policies has, for more than decade, been dominated by the imperatives of convergence. This is understandable: in the context of the EMS, it was important to bring closer the nominal performance of states.

The transition to EMU follows the same logic, which follow the convergence criteria set out in the Treaty. It has already been said that among these criteria, those aimed at public finances are problematic, because of the extent of the accumulated debt stock in some countries and the deterioration of the economic situation. On the basis of information available at mid-term 1993, it is clear that for a majority of countries a strict application of criteria for public finances would delay participation in the monetary union beyond the end of the decade, even if these countries implemented, with the costs that this entails, vigorous policies to correct their public accounts. The credibility of the whole process envisaged by the treaty would be seriously affected. That is why it is important to clarify, as soon as possible, the interpretation of public finance criteria, retaining for the public deficit the more realistic goal of a sufficient primary surplus to put the debt ratio on a robust downward path to the vagaries of growth and interest rates, and taking into account the economic situation in the judgment focused on the public deficit. It is also necessary to prevent a situation where, in a poor growth environment, several large countries would budget for correction policies simultaneously without taking into account their external impacts. In such a case of lack of coordination, the recovery could be penalised. The Treaty provides, by Article 103 (which will be applicable upon ratification), the instrument of taking into account the international dimension of the convergence process. It is important that it be makes use of it.

The question arises more generally whether the treaty satisfactorily organises the coordination of macroeconomic policies in the field of Community. Some authors (Melitz (1991), Buiter et al (1993)) argue that such coordination is useless: for them, the external effects of are low because of the limited integration of goods markets and adverse effects on the markets for goods and capital in a fixed exchange rate regime. These arguments of an empirical nature are subject to debate: according to the models, the external effects of fiscal policies vary significantly (Whitley (1991)). According to some studies, they can be strong enough to call for budget coordination (Team Mimosa (1989)). Moreover, as noted by Cohen and Wyplosz (1989), the monetary union policy creates a new externality between the participating countries. To the extent that they are not indifferent to the external balance of the union taken as a whole (for example because of the imperfect substitutability of assets and the resulting exchange rate effects), they can share an objective of external balance. In this case, coordination of budgetary policies would be imposed. Article 103 of the Treaty provides that Heads of State and Government fix large economic policy guidelines. However, there is interest in strengthening coordination procedures by setting up diagnostic devices, to alert and to follow up.

**French Economic Policy and Emu**

Although the combination of the currency crisis and the deterioration of (with the resulting deterioration of public finances) in European countries) casts doubt on the realisation of the monetary union policy, and leads to the anticipation that it will not be implemented by 1997, one of the options provided for by the Treaty, the single currency project is the objective of French economic policy.[8] It is therefore important to consider what will be the implications of participation in the single currency and how French economic policy can be prepared during the period covered by the 11th Plan. At the cost of a certain simplification, it is enlightening to distinguish implications of EMU for the stabilisation of short-medium term fluctuations and those that affect growth in the medium term. In the first case, it is necessary to examine the role of economic policies in response to temporary disruptions. In the second, it is a question of examining the paths to growth in the medium term.

**Economic stabilisation policies**

Monetary union will not eliminate the asymmetries between countries of the European Community, whether from shocks affecting them or from structures and behaviors that determine responses to common shocks. As has been said, the available body of work indicates that symmetric shocks are rather predominant in the disturbances affecting countries of the European Community. However, asymmetries will remain because of the differences in structure, institutions and behavior among national economies. Some studies even argue that they will increase in the long term because of the geographical specialisation phenomena that the single market will bring about (Eichengreen (1993)).

In a world characterised by nominal rigidities, markets do not do not adjust instantly. A decline in demand for goods translates into loss of income and unemployment. In a monetary union, a country cannot use devaluation to accelerate the adjustment of real wages. This is the main cost of monetary union. This cost is offset by the disappearance of the current balance constraint vis-à-vis other members of the union. The intensity of the external equilibrium constraint is already subject to discussion about the freedom of movement of capital (Cooper (1987)).[9]

In monetary union, the balance of payments current account vis-à-vis EMU will cease to be a constraint. On the other hand, there will still be a solvency constraint on the national economy, the nation, like any borrowing agent, must be able to pay off debts and repay them when due. A current account deficit, resulting for example from a funded capital accumulation by other countries of the Community, should not be of concern, when competitive production capacities are created guaranteeing future repayments.

---

[8] The option of an acceleration leading to the formation of a restricted monetary union is not retained here. See on this subject the report of the Workshop (Pisani-Ferry and Bismut (1993)).
[9] See also the important literature on the works of Feldstein and Horioka (1987).

If it is admitted that the private sector remains solvent (what a good functioning of the markets should ensure), the national solvency constraint is reduced to a solvency problem of general government. To suppose the excessive deficit criteria are respected, the external balance may depart from equilibrium for a long time, without movements of savings within the Community (Italianer and Pisani Ferry (1993)). In other words, the financial aspect of an external debt will be completely separate from the monetary aspects attached to it as long as the currencies remain separate and at the risk of default of the borrower can be superimposed or be substituted by a currency risk. It is likely that the single currency will lead to a further reduction of the correlation between national saving and national investment. The monetary union should completely free the Member States of the external equilibrium constraint in the short term, and better allocation of savings within the Community. However, the external balance of the Community vis-à-vis the rest of the world will not disappear and, in case of excessive imbalance, each country will have to lend attention to its contribution to the overall balance.

Thus, if we stick to a 'tinbergennian' perspective, the participation in a monetary union implies vis-à-vis the other members of the Union to both the loss of an instrument (the monetary or exchange instrument) and the loss objective (the external balance). In reality, one does not compensate exactly the other. In a random universe, the loss of an instrument leads to a loss of control capacity over the economy (Wyplosz (1992)). This loss can lead to stronger cyclical fluctuations and leads to the search for other instruments for stabilisation purposes. The gain of loosening the current balance constraint should not be overestimated. In a free movement of capital regime, this constraint is already no longer instantaneous; an intertemporal solvency constraint will subsist in a monetary union. On the other hand, the monetary union will, above all, release the economies from adverse effects in short-term capital movements.

In case of asymmetric shocks, neither labour mobility nor budget can be used as partial absorption mechanisms as happens in the United States and Canada. As a result of the many research projects brought about by the prospect of European Monetary Union, inter-regional adjustments in a union's monetary structures such as those in the United States and Canada are better known: labour mobility is an important channel of adjustment, while wage flexibility plays a very limited role (Blanchard and Katz (1992)); through the automatic interplay of taxation and transfers, the federal budget absorbs around 20% of shocks on the primary revenue of states (Sachs and Sala-y-Martin (1992), von Hagen (1991), Goodhart and Smith (1992), Bayoumi and Masson (1993), Italianer and Pisani-Ferry (1992)).

Consideration could be given to providing the European Community budget with a specific mechanism to absorb some of the transitional fluctuations in Member States income. Such a mechanism would make it possible to partially overcome the lack of labour mobility and thus strengthen the robustness of EMU in the face of shocks. In partial compensation for the abandonment of the exchange rate instrument, it would provide countries engaged in monetary union with a form of insurance. It can be shown that the budgetary cost would be very low and that the degree of stabilisation would be close to that provided by the American federal budget (Italianer and Pisani-Ferry (1992)). The potential effectiveness of such a mechanism is however the subject of discussion (Mélitz and Vori (1993)). The Treaty on the European Union did not adopt this approach.[10] It will therefore be necessary rely on national response capabilities and compensate for the lack of mobility through the use of fiscal policy and greater flexibility of wages.

---

[10] The Treaty only contemplates the possibility of exceptional financial aid (Article 103A).

Will the monetary union affect the effectiveness of fiscal policy? On the one hand, the single currency should make the policy more effective because of the dilution of the eviction and appraisal effects that follow a fiscal expansion conducted in a single country under flexible foreign exchange rates. On the other hand, in the opposite direction, economic integration increases 'leakage' that limits the Keynesian impact of a fiscal stimulus, but it is thought that this second effect will take longer to appear and should remain less important for an economy the size of France.[11] If it is clear that fiscal policy is not an exact substitute for the exchange rate, monetary union is therefore likely to give an increased role to the fiscal stabilisation policy. It will therefore be necessary to release budgetary margins to let the automatic stabilisers play without jeopardising the medium-term objectives of fiscal policy.

A key role, however, will be the flexibility of prices and wages which will be for all countries a fundamental mechanism of adjustment to shocks. In the absence of nominal flexibility of prices and wages, the adjustment will do by quantitative rationing, that is, by bankruptcies and unemployment. Instead of governments adapting wage and social behaviour to drive economic policy, in a monetary union, monetary policy will be given and social behaviour will have to adapt to it (Boyer (1993)).

However, this constraint will not mean the prevalence of a single model of social relations: to a large extent, each country may, depending on its preferences and its system of social relations, choose channels by which they will exercise this necessary flexibility. In terms of wages, this flexibility can take several alternative forms such as profit-sharing type mechanisms or wage indexing, between which a choice is possible. Moreover, it is not excluded that EMU lead to renewed use of income's policy, for example through the use of tax incentive instruments.

**Growth and employment**

Having transferred the use of the monetary policy instruments to Community level, and furthermore reserved the budgetary instrument for the purpose of stabilisation, the governments of the Member States will no longer have macroeconomic policy instruments to support growth in the medium term. They will not be able to rely on repeated devaluations nor drive policies on real interest rates or to jeopardise the equilibrium of public finances. In fact, these constraints result less from EMU than from decompartmentalisation of the financial markets which intervened in the eighties. In France, the renunciation of repeated devaluations preceded the formation of the single currency project. Many European countries have undertaken adjustment of their public finances well before the criteria for Maastricht. The macroeconomic disciplines of EMU will therefore not differ basically to those which the economies have gradually joined at the end of the experiments of the last twenty years. But they will impose themselves with even greater force on the countries of the European Community. Is that to say that they will be constantly forced to keep pace with the growth of their partners or that of the European Community as a whole?

---

[11] Fiscal policy has already lost most of its national effectiveness in small, very open economies like Belgium.

Stronger growth will not be sustainable in the long term by price competitiveness gained by the moderation of prices and wages leading to lower inflation than the European Community average. Such a policy of 'competitive disinflation'[12] is adapted to an adjustment process in which a country must strive to regain competitiveness price eroded by disinflation while respecting the constraint of fixed exchange rates: such was the case of France in recent years (Blanchard and Muet (1993)). But it would be dangerous for such strategies to be pursued in monetary union regime, for three reasons.

Firstly, he Treaty entrusts the Central Bank with the objective of price stability, it is redundant that the same objective should be pursued simultaneously by the national authorities, for example by means of an income policy. Differences in relative prices will only reflect the appreciation or depreciation of the exchange rate to the real needs of adjustment to asymmetric shocks, as in the case of federations (Poloz (1990), De Grauwe (1992)). If each country were to set a relative inflation target, a fundamental balancing mechanism characteristic of currency unions would be blocked.

Secondly, the setting of an inflation differential target can only be interpreted as an adjustment process otherwise it would be a non-cooperative policy to achieve competitiveness gains through macroeconomic policy and not by proper management of the companies. This is a zero sum game (or even negative): if all countries were to set an inflation target less than one point above average, it would be logically impossible for them to succeed and the only result would be either high unemployment or too lax monetary policy, in the event that the ECB seeks to compensate for the excess austerity of national policies.

Finally, in a monetary union regime, lower inflation is not accompanied by any decrease in nominal interest rates and therefore an equivalent increase in the ex-post real interest rate. This is the big difference to a situation of disinflation in fixed exchange rates, in which the control of prices allows a reduction in depreciation expectations and risk premiums embedded in interest rates.

EMU will not, however, require governments to passively follow the average evolution. It will instead lead them to focus on micro-economic competitiveness and growth, and to stimulate the national performance through supply-side policies and the search for social consensus. The ability of an economy to support a growth differential depends on its ability to derive non-price competitiveness gains by investment efforts, the quality of its allocation and the extension of the range of varieties offered. In this perspective, the effectiveness of the State in production of public goods, the adaptation of tax levy systems and social benefits for an economic efficiency and social equity, the quality of public infrastructure, the impact of research spending, the performance of the education system will be for each country key drivers of competitiveness and growth in the medium term. This is why the prospect of EMU draws upon all of the consequences of recent research on the determinants of growth and the impact of public policies on the growth of economies.

Much the same goes for employment. Growth of labour resources of the European economies over the last ten or twenty next few years will not be homogeneous, and international mobility will remain a marginal adjustment channel. These differences can only be compensated partially by the growth differentials, and all the more so if the stock of unemployed is taken into account. EMU may create favourable conditions for European growth, but it will be up to each country to make use of the instruments at its disposal to reverse the spontaneous trend in labour productivity.

---

[12]Competitive disinflation has been the subject of many interpretations. We by this means a policy of disinflation aiming at a certain gap inflation, positive or negative, with a partner country (or a group of countries)

**How to prepare for a change in monetary regime?**

The French economy is preparing to change its political economic regime. From the first oil shock to the end of the eighties, concepts and objectives have been dominated by the notion of external constraint. 'Competitive disinflation' itself was part of this culture, since it found a justification in the volume gains that allowed (until the devaluations of autumn 1992) the progress of price competitiveness, as much or more than on the benefits of price stability themselves. The Plan played a leading role in the analysis and pedagogy of the disciplines of the open economy. In the coming years, this constraint will complete to fade, while it will be substituted for a flexibility constraint of prices and wages.

This perspective is an extension of the efforts undertaken since more than ten years. But if the success of disinflation is indicative of the scale of these efforts, econometric research does not confirm that price and wage behaviour have changed in an obvious way (Blanchard and Mute (1993), Egebo and Englander (1992)). The coming years must therefore be used to continue the preparation of the French economy for the new rules of the game. It will be necessary to tackle the search for greater responsiveness of prices and wages to internal or external disturbances and strengthening of the micro-economic instruments of stimulation for growth. Economic agents will have to learn to listen to the signals that the Central Bank will send them. At the same time, the framework of reference for macroeconomic policy will need to be developed, with the completion of the adjustment symbolised by competitive disinflation and the renewal of hierarchy objectives. Finally, it will be necessary to develop a pedagogy of the disciplines associated with EMU and organise a social dialogue.